\documentclass[a4paper,english,titlepage,12pt]{article}

\usepackage{babel}\usepackage{amsmath,amssymb,amsthm}
\usepackage[numbers,compress]{natbib}
\usepackage{graphicx}
\usepackage{longtable}
\usepackage{float}
\usepackage{titlesec}
\usepackage{caption}

\makeatletter
\def\@xfootnote[#1]{%
  \protected@xdef\@thefnmark{#1}%
  \@footnotemark\@footnotetext}
\makeatother

\newcommand{\new}{\text{\textit{\textbf{new}}}}
\newcommand{\rop}{\mathcal{R}}
\newcommand{\ffdeg}{\mathrm{Deg}}
\newcommand{\forests}{\mathfrak{F}}
\newcommand{\infragr}{\mathfrak{I}}

\newcommand{\cbigz}{C_{\text{bigZ}}}
\newcommand{\cbigf}{C_{\text{bigF}}}
\newcommand{\csubse}{C_{\text{subSE}}}
\newcommand{\csubi}{C_{\text{subI}}}
\newcommand{\csubo}{C_{\text{subO}}}
\newcommand{\cadd}{C_{\text{add}}}

\renewcommand{\thesection}{\Roman{section}}

\makeatletter
\renewcommand{\p@subsection}{\thesection.}
\makeatother
\titlelabel{\thetitle.\ }

\begin{document}

\begin{center}
\LARGE{\textbf{Calculating the 5-loop QED contribution to the electron anomalous magnetic moment: graphs without lepton loops}}
\end{center}
\begin{center}
\large{Sergey Volkov\footnote[*]{E-mail: \texttt{volkoff\char`_sergey@mail.ru, sergey.volkov.1811@gmail.com}}}
\\ \ \\ \emph{\small{SINP MSU, Moscow, Russia \\ DLNP JINR, Dubna, Russia}}
\end{center}

\small{This paper describes a computation of a part of the QED contribution to the electron anomalous magnetic moment that was performed by the author with the help of a supercomputer. The computed part includes all 5-loop QED Feynman graphs without lepton loops. The calculation has led to the result $A_1^{(10)}[\text{no lepton loops}]=6.793(90)$ that is slightly different than the value $7.668(159)$ presented by T. Aoyama, T. Kinoshita, and M. Nio in 2018. The discrepancy is about $4.8\sigma$. The computation gives the first independent check for that value. A shift in the fine-structure constant prediction is revealed in the paper. The developed calculation method is based on (a) a subtraction procedure for removing all ultraviolet and infrared divergences in Feynman parametric space before integration; (b) a nonadaptive Monte Carlo integration that uses the probability density functions that are constructed for each Feynman graph individually using its combinatorial structure. The method is described briefly in the paper (with the corresponding references to the previous papers). The values for the contributions of nine gauge-invariant classes splitting the whole set are presented in the paper. Moreover, the whole set of all 5-loop graphs without lepton loops is split into 807 subsets for comparison (in the future) of the calculated values with the values obtained by another methods. These detailed results are presented in the supplemental materials. Also, the supplemental materials contain the contribution values for each of 3213 individual Feynman graphs. An ``oscillating'' nature of these values is discussed. A realization of the numerical integration on the graphics accelerator NVidia Tesla V100 (as a part of the supercomputer ``Govorun'' from JINR, Dubna) is described with technical details such as pseudorandom generators, calculation speed, code sizes and structure, prevention of round-off errors and overflows, etc.
} \normalsize

\section{INTRODUCTION}

The most precise measurement of the electron anomalous magnetic moment (AMM) gave the result
\begin{equation}\label{eq_ae_meas}
a_e[\text{expt.}]=0.00115965218073(28).
\end{equation}
This result was presented by Gabrielse research group at Harvard in Ref.~\cite{experiment}.
All theoretical predictions for $a_e$ must satisfy this ``quality standard'' for the precision. The ``mainstream'' Standard Model prediction uses the following expression:
$$
a_e=a_e(\text{QED})+a_e(\text{hadronic})+a_e(\text{electroweak}),
$$
$$
a_e(\text{QED})=\sum_{n\geq 1} \left(\frac{\alpha}{\pi}\right)^n
a_e^{2n},
$$
$$
a_e^{2n}=A_1^{(2n)}+A_2^{(2n)}(m_e/m_{\mu})+A_2^{(2n)}(m_e/m_{\tau})+A_3^{(2n)}(m_e/m_{\mu},m_e/m_{\tau}),
$$
where $m_e,m_{\mu},m_{\tau}$ are the masses of the electron, muon and tau-lepton, respectively. The universal QED terms $A_1^{(2n)}(\alpha/\pi)^n$ form the most significant contribution to the value. The coefficient values
$$
A_1^{(2)}=0.5,\quad A_1^{(4)}=-0.328478965579\ldots
$$
were presented in Refs. \cite{schwinger1,schwinger2} and Refs. \cite{analyt2_p,analyt2_z}, respectively. The value of $A_1^{(6)}$ was being calculated in 1970-x by different groups of scientists using numerical integration; see Refs. ~\cite{carrollyao,carroll}, ~\cite{levinewright}, ~\cite{kinoshita_6}. The most accurate value $A_1^{(6)}=1.195\pm 0.026$ for that era was obtained in 1974 by T. Kinoshita and P. Cvitanovi\'{c}. The uncertainty is caused by the statistical error of the Monte Carlo integration. A work of analytical calculation of $A_1^{(6)}$ with the help of computers was started at the same time. The final value 
$$
A_1^{(6)}=1.181241456\ldots
$$
was obtained by S. Laporta and E. Remiddi in 1996; see Ref. ~\cite{analyt3}. That value was a product of efforts of many researchers; see, for example, Refs. ~\cite{analyt_mi,analyt_b2,analyt_b3,analyt_b1,analyt_b4,analyt_b,analyt_i,analyt_j,analyt_g,analyt_e,analyt_h,analyt_d,analyt_c,analyt_ll,laporta_1993,analyt_f}. First numerical estimations for $A_1^{(8)}$ were obtained by T. Kinoshita and W. B. Lindquist in 1981 and published in Ref. ~\cite{kinoshita_8_first}. The most accurate value presented by T. Kinoshita's team
$$
A_1^{(8)}=-1.91298(84)
$$
was published in 2015 in Ref. ~\cite{kinoshita_8_last}. That value was obtained by Monte Carlo integration. S. Laporta's semianalytical result
$$
A_1^{(8)}=-1.9122457649\ldots
$$
was obtained in 2017 and published in Ref. ~\cite{laporta_8}. These two calculations of $A_1^{(8)}$ are in good agreement as well as another independent calculations of this value from Refs. ~\cite{smirnov_amm,rappl}, and for Feynman graphs without lepton loops from Ref. ~\cite{volkov_gpu}.

The full calculation of $A_1^{(10)}$ was performed only by T. Kinoshita's team using Monte Carlo integration. The most precise value was obtained in 2019 by T. Aoyama, T. Kinoshita, M. Nio and was published in Ref. ~\cite{kinoshita_atoms}:
\begin{equation}\label{eq_kinoshita_10}
A_1^{(10)}[\text{AKN}]=6.737(159).
\end{equation}
A special place is occupied by the contribution of Feynman graphs without lepton loops to $A_1^{(10)}$. This set contains 3213 Feynman graphs\footnote{Graphs that are obtained from each other by changing arrow directions are regarded as one.} and forms a gauge-invariant class. This contribution is the most complicated one for both Monte Carlo integration and analytical calculations. For example, the uncertainty in (\ref{eq_kinoshita_10}) is entirely determined by that contribution. Also, it is the contribution that suffered the most from found mistakes and corrections; see Ref. ~\cite{kinoshita_10_corr}. The value
\begin{equation}\label{eq_kinoshita_10_nolepton}
A_1^{(10)}[\text{no lepton loops, AKN}]=7.668(159).
\end{equation}
can be obtained by using (\ref{eq_kinoshita_10}) and the value of the remaining part that can be extracted from Ref. ~\cite{kinoshita_10_corr}. By 2019, there was no independent calculations of $A_1^{(10)}[\text{no lepton loops}]$.

We recalculated this contribution with the help of the supercomputer ``Govorun'' (JINR, Dubna, Russia). 40000 GPU-hours of Monte Carlo integration on NVidia Tesla V100 that were spread over several months have led to the result
\begin{equation}\label{eq_nolepton}
A_1^{(10)}[\text{no lepton loops, Volkov}]=6.793(90),
\end{equation}
where the uncertainty corresponds to $1\sigma$ limits. It is in good agreement with the preliminary value $6.782(113)$ published in Ref. ~\cite{volkov_acat}. The descrepancy between this result and (\ref{eq_kinoshita_10_nolepton}) is approximately $4.8\sigma$. This means that the values are probably different. The reason of this difference is unknown. Sec. \ref{sec_res_tech} contains some considerations about reliability of the result. In addition, it is important that this result can be checked by parts; see the detailed explanation in Sec. \ref{sec_res_tech}.

Combining (\ref{eq_nolepton}) with the value of the residual part of $A_1^{(10)}$ from Ref. ~\cite{kinoshita_10_corr}, we obtain
\begin{equation}\label{eq_volkov_akn_10}
A_1^{(10)}[\text{Volkov+AKN}]=5.862(90).
\end{equation}
Taking the known and double-checked values for $A_2^{(2n)}$, $n\leq 5$, $A_3^{(2n)}$, $n\leq 4$, $a_e(\text{hadronic})+a_e(\text{electroweak})$ (see a review in Ref. ~\cite{kinoshita_atoms}) and the measured value of $\alpha$ from Ref. ~\cite{alpha_cesium} based on a measurement of the cesium atom mass relative to the Planck constant
\begin{equation}\label{eq_alpha_cesium}
\alpha^{-1}(\text{Cs})=137.035999046(27),
\end{equation}
we obtain
$$
a_e[\text{theory},\alpha(\text{Cs}),\text{Volkov}]=0.001159652181547(6)(12)(229),
$$
where the first uncertainty comes from (\ref{eq_nolepton}), the second one from the hadronic and electroweak corrections, and the last one from the uncertainty of $\alpha$. The usage of (\ref{eq_kinoshita_10}) will give
$$
a_e[\text{theory}, \alpha(\text{Cs}),\text{AKN}]=0.001159652181606(11)(12)(229)
$$
instead. If we will use the $a_e$ prediction with (\ref{eq_volkov_akn_10}) and the measured value (\ref{eq_ae_meas}) for improving $\alpha$, we obtain
\begin{equation}\label{alpha_ae_volkov}
\alpha^{-1}[a_e,\text{Volkov}]=137.0359991427(7)(14)(331),
\end{equation}
where the uncertainties come from (\ref{eq_nolepton}), the hadronic and electroweak corrections, (\ref{eq_ae_meas}), correspondingly. The discrepancy with (\ref{eq_alpha_cesium}) is approximately $2.27\sigma$. The corresponding value obtained from (\ref{eq_kinoshita_10}) is
\begin{equation}\label{alpha_ae_akn}
\alpha^{-1}[a_e, \text{AKN}]=137.0359991496(13)(14)(330)
\end{equation}
with the discrepancy $2.43\sigma$ relative to (\ref{eq_alpha_cesium}). If we take 
\begin{equation}\label{eq_alpha_rubidium}
\alpha^{-1}(\text{Rb})=137.035998996(85)
\end{equation}
obtained from the measurement of the rubidium atom mass relative to the Planck constant (Ref. ~\cite{alpha_rubidium}) combined with the improved values of some constants from CODATA-2014 (Ref. ~\cite{codata_2014}), we obtain
$$
a_e[\text{theory},\alpha(\text{Rb}),\text{Volkov}]=0.001159652181969(6)(12)(720),
$$
$$
a_e[\text{theory}, \alpha(\text{Rb}),\text{AKN}]=0.001159652182037(11)(12)(720).
$$
The values (\ref{alpha_ae_volkov}) and (\ref{alpha_ae_akn}) have the discrepancies $1.61\sigma$ and $1.69\sigma$ relative to (\ref{eq_alpha_rubidium}). This means that the discrepancy between (\ref{eq_nolepton}) and (\ref{eq_kinoshita_10_nolepton}) affects $\alpha$ and $a_e$ slightly. However, this discrepancy can become significant in the future, when the precision of the measurements will be increased. Also, if both calculations have mistakes, then this can be sensible even at the current level of precision. Thus, an additional independent calculation is required.

There is no universal method that makes it possible to calculate 5-loop QED contributions in a realistic time frame. Firstly, the existing universal IR divergence control methods like those that are based on the dimensional regularization lead to enormous amounts of symbolic manipulations. And secondly, the universal integration routines demonstrate a very slow convergence on the obtained integrals. 

To make the 5-loop calculations practically feasible it is required to remove all ultraviolet (UV) and infrared (IR) divergences before integration and to avoid any $\varepsilon$-like regularizations. All UV divergences in Feynman integrals can be removed by the direct subtraction on the mass shell using a forestlike formula like Zimmermann's forest formula\footnote{The Zimmermann forest formula was first published in Ref. ~\cite{scherbina} and Ref. ~\cite{zavialovstepanov}. However, the historic name is connected with Ref. ~\cite{zimmerman}.}. However, an analogous method for removing IR divergences has not been invented yet. The anomalous magnetic moment is free from IR divergences: the IR divergences corresponding to soft virtual photons are compensated by the IR divergences connected with the on-shell renormalization; see notes in Ref. ~\cite{volkov_prd}. But unfortunately, direct methods lead to an emergence of IR divergences in individual Feynman graphs. Different authors use different homemade divergence subtraction procedures that work in some cases; see Refs. \cite{carrollyao,levinewright,kinoshita_infrared,kinoshita_atoms}. A relatively simple subtraction procedure giving finite Feynman parametric integrals was developed for our calculations. It was presented firstly in Ref. ~\cite{volkov_2015} and is briefly described in Sec. \ref{sec_subtraction}.

The 5-loop calculations lead to Feynman parametric integrals with 13 variables. At this time, the only way to evaluate such integrals numerically is to use Monte Carlo integration. Unfortunately, Feynman parametric integrands after divergence subtraction are unbounded and have a very complicated asymptotic behavior near boundaries. The universal adaptive Monte Carlo integration routines like VEGAS can, in principle, work with unbounded functions and functions having a steep landscape. However, these routines are suited for functions with a certain shape. This becomes critical for large numbers of variables. For example, VEGAS uses the probability density functions of the form
$$
f_1(x_1)\cdot f_2(x_2)\cdot\ldots\cdot f_n(x_n)
$$
and tries to fit the functions $f_j$ to make the convergence as fast as possible\footnote{The Monte Carlo integration error usually behaves as $\sigma\sim C/\sqrt{N}$, where $N$ is the number of samples. However, it is very important to make $C$ as small as possible.}. Unfortunately, this approximation does not work fine for Feynman parametric integrals with large numbers of variables. A nonadaptive\footnote{except the inter-graph adaptivity described in Sec. \ref{subsec_monte_carlo_details} and the adjustment of six constants (\ref{eq_mc_settings}) that was performed once for the 4-loop graphs} method that uses some a priori knowledge about the Feynman parametric integrands behavior was developed for our calculations. The method that is briefly described in Sec. \ref{sec_monte_carlo} works only for graphs without lepton loops. The first version of this method was presented in Ref. ~\cite{volkov_prd}.

The developed Monte Carlo integration method allows us to reduce the needed number of samples substantially. However, in the 5-loop case, for evaluating 3213 Feynman graphs a supercomputer is still required. Modern graphics processors (GPUs) are more suitable for performing many uniform sequences of arithmetic operations in parallel than usual processors. The Monte Carlo integration was performed on GPUs NVidia Tesla V100 as a part of the supercomputer\footnote{The GPU part of the supercomputer ``Govorun'' has 40 GPUs NVidia Tesla V100. The peak performance of the GPU part is 300 TFlops for double precision. The peak performance of the whole supercomputer (including the CPU part) is 500 TFlops.} ``Govorun'' from JINR (Dubna, Russia). The realization is described in Sec. \ref{sec_realization} with some programming details. Sec. \ref{sec_res_tech} contains the results of the calculations, a discussion about these results, the description of the supplemental materials, and some technical information about the computation including the GPU performance, arithmetic precision statistics and so on.

\section{DIVERGENCE ELIMINATION}\label{sec_subtraction}

The developed subtraction procedure is based on a forest formula with linear operators that are applied to the Feynman amplitudes of UV divergent subgraphs. This is similar to the Zimmermann forest formula. The difference is only in the choice of the linear operators used and in the way of combining them. Let us recapitulate the advantages of the developed procedure:
\begin{itemize}
\item The procedure is fully automated for any order of the perturbation series\footnote{The method must work for all Feynman graphs contributing to $A_1^{(2n)}$ including the ones containing lepton loops; see Ref. ~\cite{volkov_2015}. However, a rigorous mathematical proof for this fact is not developed even for graphs without lepton loops.}.
\item The method is beautiful and is relatively simple for realization on computers.
\item The subtraction is equivalent to the on-shell renormalization: for obtaining the final result we should only sum up the contributions of all Feynman graphs after subtraction. Thus, no residual renormalizations are required.
\item Feynman parameters can be used directly, without any additional tricks.
\end{itemize}

There are the following types of UV-divergent subgraphs\footnote{We consider only such subgraphs that are strongly connected and contain all lines that join the vertexes of the given subgraph.} in QED Feynman graphs without lepton loops: \emph{electron self-energy subgraphs} ($N_e=2,N_{\gamma}=0$)
and \emph{vertexlike} subgraphs ($N_e=2,N_{\gamma}=1$), where by $N_e$ and $N_{\gamma}$ we denote the number of external electron and photon lines in the subgraph.

Two subgraphs are said to overlap if they are not contained one
inside the other, and the intersection of their sets of lines is not empty.

A set of subgraphs of a graph is called a \emph{forest} if any two
elements of this set do not overlap.

For a vertexlike graph $G$ by $\forests[G]$ we denote the set of all
forests $F$ that consist of UV-divergent subgraphs of $G$ and
satisfy the condition $G\in F$. By $\infragr[G]$ we denote the
set of all vertexlike subgraphs $G'$ of $G$ such that $G'$ contains
the vertex that is incident\footnote{We say that a line $l$ and a
vertex $v$ are \emph{incident} if $v$ is one of the endpoints of
$l$.} to the external photon line of $G$.\footnote{In particular,
$G\in \infragr[G]$.}

We work in the system of units, in which $\hbar=c=1$, the
factors of $4\pi$ appear in the fine-structure constant:
$\alpha=e^2/(4\pi)$, the tensor $g_{\mu\nu}$ is defined by
$$
g_{\mu\nu}=g^{\mu\nu}=\left(\begin{matrix}1 & 0 & 0 & 0 \\ 0 & -1 &
0 & 0 \\ 0 & 0 & -1 & 0 \\ 0 & 0 & 0 & -1 \end{matrix}\right),
$$
the Dirac gamma-matrices satisfy the condition
$\gamma^{\mu}\gamma^{\nu}+\gamma^{\nu}\gamma^{\mu}=2g^{\mu\nu}$.

The following linear operators are used for the subtraction:
\begin{enumerate}
\item $A$ is the projector of the AMM. This
operator is applied to the Feynman amplitudes of vertexlike
subgraphs. See the definition in Refs. ~\cite{volkov_2015,volkov_prd}.
\item The definition of the operator $U$ depends on the type of
UV-divergent subgraph to which the operator is applied:
\begin{itemize}
\item
If $\Sigma(p)$ is the Feynman amplitude that corresponds to an
electron self-energy subgraph,
$$
\Sigma(p)=u(p^2)+v(p^2)\hat{p},
$$
then, by definition\footnote{Note that it differs from the standard
on-shell renormalization.},
$$
U\Sigma(p) = u(m^2)+v(m^2)\hat{p},
$$
where $m$ is the mass of the electron, $\hat{p}=p^{\mu}\gamma_{\mu}$.
\item If $\Gamma_{\mu}(p,q)$ is the Feynman amplitude corresponding to a vertexlike subgraph,
\begin{equation}\label{eq_gamma_general_q0}
\Gamma_{\mu}(p,0)=a(p^2)\gamma_{\mu} + b(p^2)p_{\mu} +
c(p^2)\hat{p}p_{\mu}+d(p^2)(\hat{p}\gamma_{\mu}-\gamma_{\mu}\hat{p}),
\end{equation}
then, by definition,
$$
U\Gamma_{\mu}=a(m^2) \gamma_{\mu}.
$$
\end{itemize}
\item $L$ is the operator that is used in the standard subtractive on-shell renormalization
of vertexlike subgraphs. If $\Gamma_{\mu}(p,q)$ is the Feynman
amplitude that corresponds to a vertexlike subgraph,
(\ref{eq_gamma_general_q0}) is satisfied, then, by definition,
$$
L\Gamma_{\mu}=[a(m^2)+mb(m^2)+m^2c(m^2)]\gamma_{\mu}.
$$
\end{enumerate}

Let $f_G$ be the unrenormalized Feynman amplitude that corresponds
to a vertexlike graph $G$. Let us write the symbolic definition
$$
\tilde{f}_G=\rop^{\new}_G f_G,
$$
where
$$
\rop^{\new}_G=\sum_{\substack{F=\{G_1,\ldots,G_n\}\in \forests[G] \\
G'\in \infragr[G]\cap F}}(-1)^{n-1}M^{G'}_{G_1}M^{G'}_{G_2}\ldots
M^{G'}_{G_n},
$$
$$
M^{G'}_{G''}=\begin{cases}A_{G'},\text{ if }G'=G'', \\
U_{G''},\text{ if }G''\notin \infragr[G]\text{, or }G''\varsubsetneq
G',
\\ L_{G''},\text{ if }G''\in \infragr[G], G'\varsubsetneq G'', G''\neq
G,
\\ (L_{G''}-U_{G''}),\text{ if }G''=G, G'\neq G.\end{cases}
$$
In this notation, the subscript of an operator symbol denotes the
subgraph to which this operator is applied.

The coefficient before $\gamma_{\mu}$ in $\tilde{f}_G$ is the
contribution of $G$ to $a_e$.

For example, for the graph $G$ from FIG. \ref{fig_graph_eia_fail} we will have the following operator expression:
\begin{equation}\label{eq_example_expression}
[A_G(1-U_{bcdefghij})-(L_G-U_G)A_{bcdefghij}](1-U_{cd})(1-U_{fghi})(1-U_{fgh}-U_{ghi}).
\end{equation}
Here the subscripts mean the subgraphs to which the operators are applied (denoted by the enumeration of the vertexes). The expression means that we should remove brackets, and for each term we should transform the Feynman amplitudes of the subgraphs using the corresponding operators from the inner subgraphs to the outer ones. The transformation is applied in Feynman parametric space before integration. This can be explained easy using the approach to Feynman parameters based on the transferring from Schwinger parameters; see Ref. ~\cite{volkov_2015}.

The operators $U$ are designed for removing UV divergences in the way similar to the Zimmermann forest formula and Bogoliubov's R-operation. In contrast to the usual for QED operator $L$ the operators $U$ do not generate additional IR divergences. The multiplier in the square brackets in (\ref{eq_example_expression}) corresponds to elimination of the IR divergences that correspond to soft virtual photons on the external electron lines and the UV divergences connected with the subgraphs to which the operators are applied. Also, the ``overall'' UV and IR divergences are removed by the magnetic moment projector $A$ as well as it works in the 1-loop case; see ~\cite{kinoshita_infrared} and ~\cite{volkov_2015}. It is important that the operator $U$ applied to self-energy subgraphs extracts the self-mass part completely. This allows us to avoid IR divergences of power type; see Discussion in Ref. ~\cite{volkov_2015}. The cancellation of divergences is described in detail\footnote{although not completely rigorously} in terms of Feynman parameters in Ref. ~\cite{volkov_2015}; see also additional comments in Ref. ~\cite{volkov_gpu}.

The equivalence of the subtraction procedure and the direct subtraction on the mass shell is proved in a combinatorial way in Ref. ~\cite{volkov_2015}, Appendix B. For proving this equivalence we use the fact that the operator $U$ preserves the Ward identity; see Ref. ~\cite{volkov_2015}. It is easy to see this equivalence in the 2-loop case; see Section 3 of Ref. ~\cite{volkov_2015}. Let us note that we do not use the operator of QED on-shell renormalization of electron self-energy subgraphs; the Ward identity helps us in this case too. For a detailed explanation of the developed method, see Ref. ~\cite{volkov_2015} and some additional explanations in Refs. ~\cite{volkov_prd,volkov_gpu}.

\section{MONTE CARLO INTEGRATION}\label{sec_monte_carlo}

\subsection{Probability density functions}

After removing divergences the contribution of each Feynman graph to $A_1^{(2n)}$ is represented as an integral of the form
\begin{equation}\label{eq_feyn_integral}
\int_{z_1,\ldots,z_M>0}I(z_1,\ldots,z_M)\delta(z_1+\ldots+z_M-1)dz_1\ldots
dz_M,
\end{equation}
where $M=3n-1$ (see\footnote{We use a trick for reducing the number from $3n$ to $3n-1$; see ~\cite{volkov_prd}.}), $z_j$ are the Feynman parameters. For each graph we calculate the $(3n-2)$-dimensional integral directly; we do not use any additional reductions.

We propose to split all the integration area into the Hepp sectors (see Ref. ~\cite{hepp}) that are simply orders on the Feynman parameters:
$$
z_{j_1}\geq z_{j_2}\geq\ldots\geq z_{j_M}.
$$

We use the probability density functions of the form
\begin{equation}\label{eq_pdf_final}
g(\underline{z})=C_1g_1(\underline{z}) + C_2g_2(\underline{z}) +
C_3g_3(\underline{z}) + C_4g_4(\underline{z}),
\end{equation}
where $\underline{z}=(z_1,\ldots,z_M)$,
\begin{equation}\label{eq_pdf}
g_1(\underline{z})=C\cdot \frac{\prod_{l=2}^M \left(z_{j_l}/z_{j_{l-1}}\right)^{\ffdeg(\{ j_l,j_{l+1},\ldots,j_M\})}}{z_1\cdot z_2\cdot\ldots \cdot z_M},
\end{equation}
$C_1,C_2,C_3,C_4$ are some constants (see Sec. \ref{sec_realization}), $\ffdeg(s)$ are positive real numbers for each set $s$ of internal lines\footnote{If we use the trick for reducing the number of variables by one, we consider two electron lines that adjoin the external photon line as one line.} of the graph (except the empty and full sets), $C$ is the normalization constant defined by
$$
\int_{z_1,\ldots,z_M>0}g_1(z_1,\ldots,z_M)\delta(z_1+\ldots+z_M-1)dz_1\ldots
dz_M = 1.
$$
The stabilization functions $g_2,g_3,g_4$ are defined in Ref. ~\cite{volkov_gpu}; an additional constant $D$ is used for defining $g_3$.

Functions of the form (\ref{eq_pdf}) was first used for approximating the behavior of parametric integrals by E. Speer; see Ref. ~\cite{speer}.

The main problem in this approach is that for good Monte Carlo convergence the values $\ffdeg$ must be adjusted very accurately. Speer's lemma (Ref. ~\cite{speer}) states that in some simple cases, when we do not have UV divergent subgraphs and we do not consider the infrared behavior, we may take the ultraviolet degree of divergence (with the sign minus) of $s$ as $\ffdeg(s)$ and use (\ref{eq_pdf}) as an upper bound for $|I(\underline{z})|$. A good upper bound can play the role of a good probability density function for Monte Carlo integration; see Ref. ~\cite{volkov_prd}. However, in the real case we should use a more complicated formulas for obtaining $\ffdeg(s)$. These formulas were developed for our calculations\footnote{However, a rigorous mathematical proof that the expressions of this form can be used as upper bounds for $I(\underline{z})$ has not been obtained yet. The assurance is based on numerical experiments.}. The first version of the method was presented in Ref. ~\cite{volkov_prd}. We use an improved version from Ref. ~\cite{volkov_gpu}. The algorithm of obtaining $\ffdeg(s)$ uses six constants $\cbigf>0$, $\cbigz>0$, $\cadd$, $\csubi$, $\csubse$, $\csubo$ that should be choosed by hand. For the 5-loop case we use the same values as we used for the 4-loop, 3-loop, and 2-loop cases in Ref. ~\cite{volkov_gpu}:
\begin{equation}\label{eq_mc_settings}
\begin{array}{c}
\cbigz=0.256,\ \cbigf=0.839,\ \cadd=0.786,\ \\  \csubi=0.2,\
\csubse=0 ,\ \csubo=0.2.
\end{array}
\end{equation}
These values were obtained by numerical experiments with 4-loop graphs.
Note that some of the values $\ffdeg(s)$, obtained by the method, less than $1$ and even sometimes less than $1/3$, in contrast to integer numbers in Speer's lemma (Ref. ~\cite{speer}).

The terms $C_jg_j(\underline{z})$, $j=2,3,4$ in (\ref{eq_pdf_final}) are added for ensurance: they cannot slow down the Monte Carlo convergence speed significantly, but they can (in principle) prevent from occasional emergence of gigantic contributions of some samples; see Ref. ~\cite{volkov_gpu}.

The algorithm of fast random sample generation is described in Ref. ~\cite{volkov_prd}.

\subsection{Obtaining the value and uncertainty}\label{subsec_carlo_values}

If the random samples $\underline{z}_1,\ldots,\underline{z}_N$ are generated with the probability density function $g(\underline{z})$, then the integral value is approximated as
\begin{equation}\label{eq_integral_approx}
\frac{1}{N}\sum_{j=1}^N \frac{I(\underline{z}_j)}{g(\underline{z}_j)}.
\end{equation}
For approximating the standard deviation $\sigma$ we can use the formula
\begin{equation}\label{eq_sigma_down}
\sigma^2=\frac{\sum_{j=1}^N y_j^2}{N^2} -
\frac{\left(\sum_{j=1}^N y_j \right)^2}{N^3},
\end{equation}
where $y_j=I(\underline{z}_j)/g(\underline{z}_j)$.
However, in practice this formula often leads to an underestimation of the standard deviation. The reason is that the real $\sigma^2$ is the mean value of the right part of (\ref{eq_sigma_down}), but using (\ref{eq_sigma_down}) we will rather obtain something near the median of that value that is often less than the mean value. Taking into account this difference is especially important when we integrate unbounded functions. Because of this, we use an improved value $\sigma_{\uparrow}$ as $\sigma$ instead of (\ref{eq_sigma_down}). The algorithm of obtaining $\sigma_{\uparrow}$ based on heuristic predictions is described in Ref. ~\cite{volkov_gpu}. For the 5-loop case we use exactly the same method. The value defined by (\ref{eq_sigma_down}) we denote by $\sigma_{\downarrow}$. A large value of $\sigma_{\uparrow}/\sigma_{\downarrow}$ indicates that the obtained integral value is suspicious, but no guarantees are possible for Monte Carlo integration. We use $\sigma_{\uparrow}$ for all intervals in the paper.

\section{REALIZATION}\label{sec_realization}

\subsection{Evaluation of the integrands with GPUs}\label{subsec_realization_integrands}

The code for all 3213 integrands was generated automatically. The D programming language was used for the codegenerator; see Ref. ~\cite{dlang}. The generated code was written in C++\footnote{We did not use any substantial improvement of C++ over C like object oriented programming for the generated code. But some little improvements were used, so we must call it ``C++'', not ``C''.} with CUDA; see Ref. ~\cite{cuda}. The codegeneration took about one month on two CPU cores of a personal computer. 

Numerical subtraction of divergences under the integral sign can cause round-off errors. We use interval arithmetic (IA) for controlling them. In interval arithmetic we work not with numbers, but with intervals of numbers. NVidia GPUs support all necessary operations for the realization of interval arithmetic. However, arithmetic operations with intervals are slow, and we developed a fast modification of interval arithmetic that was called ``\emph{eliminated interval arithmetic}'' (EIA). The main idea of EIA is that in some cases we can replace a large sequence of interval arithmetic operations by the analogous sequence of operations on the centers of the intervals and estimate the radius of the final interval by a relatively simple formula. The intervals obtained by EIA are wider than the ones obtained by IA, but both of them are reliable. EIA is described in detail in Ref. ~\cite{volkov_gpu}.

The integrals for all Feynman graphs are calculated simultaneously; see Sec. \ref{subsec_monte_carlo_details}.  At the stage of inititialization, we evaluate approximately $10^8$ random points for each Feynman graph with the machine double-precision IA taking the nearest to zero point of each interval. After initialization, when we evaluate the value of $I(\underline{z})/g(\underline{z})$ from (\ref{eq_integral_approx}) at some point $\underline{z}$, we first calculate it using EIA. The obtained interval $[y^-;y^+]$ is accepted if\footnote{This criteria differs from the previous one from Ref. ~\cite{volkov_gpu}. The previous criteria was erroneous: it did not take into account that the mean value of the round-off error is not zero. However, that error did not significantly affect the result.}
\begin{equation}\label{eq_interval_accept}
y^+-y^- \leq \frac{1}{4}\sigma_{\downarrow,j} \cdot \frac{\sqrt{\sum_l (\sigma_{\downarrow,l})^2}}{\sum_l \sigma_{\downarrow,l}},
\end{equation}
where the summations go over all contributing Feynman graphs, $j$ is the number of the current graph, $\sigma_{\downarrow,l}$ is the value of $\sigma_{\downarrow}$ calculated for the integral corresponding to the graph with the number $l$. This formula guarantees that the total round-off error (summed over all graphs) does not exceed $C\sigma_{\downarrow}$ for some constant $C$. Also, it satisfies the natural demand that larger round-off errors are possible for graphs with larger $\sigma_{\downarrow,l}$. If the interval was not accepted, it is recalculated using IA with increased precisions until it is accepted: machine double precision, 128-bit-mantissa precision, 192-bit-mantissa precision, 256-bit-mantissa precision. If all precisions failed, then the contribution is supposed to be zero. EIA fails approximately on one in five samples. However, the integrand evaluation in EIA is approximately 6.5 times faster than in the double-precision IA; see Sec. \ref{sec_res_tech} and Table \ref{table_tech}. Thus, the usage of EIA significantly improves the performance.

The Monte Carlo samples are generated and performed by blocks. Each block contains approximately $10^9$ samples pertaining to a single Feynman graph. The block scheduling algorithm is described in Sec. \ref{subsec_monte_carlo_details}. The samples are processed on a GPU in 20480 parallel threads\footnote{80 blocks of 256 threads; see ~\cite{cuda}.}. Each thread processes some set of the block samples sequentially. Branching is not allowed in the execution of a code for GPU, so the samples requiring increased precision are collected and then processed in the subsequent GPU calls.

We use a handmade library for arbitrary precision arithmetic. The 128-bit-mantissa arithmetic is realized using the GPU register memory\footnote{The register memory is the fastest kind of memory in NVidia GPUs.}. The greater precisions are realized with the global GPU memory. The usage of the register memory improves the performance by approximately 10 times\footnote{However, Table \ref{table_tech} shows a more significant gap. That is because there are very few points that require 192-bit-mantissa and more precision, and the GPU parallelism can not be exploited for all its worth on these points.}. Nevertheless, the increased precision calculations occupy a considerable part of the calculation time; see Sec. \ref{sec_res_tech} and Table \ref{table_tech}.

For each integrand we generate program codes for three precisions separately: EIA, double-precision IA, and arbitrary-precision IA. This leads to a relatively large code. The total size of the integrands code is 400 GB in the not compiled form and 500 GB in the compiled form.

The calculation of some integrand values requires millions of arithmetic operations. However, both compilers and optimizers do not like big functions. We split the calculation of each integrand into several CUDA kernels\footnote{A CUDA kernel is a GPU function that is called from the CPU part; see ~\cite{cuda}.}. Each CUDA kernel contains approximately 3000 arithmetic operations for the EIA code, 2000 operations for the double-precision IA code, and 1000 operations for the arbitrary-precision IA code. The arbitrary-precision integrand code is also split into several files: approximately 50 CUDA kernels per file. The choice of the function sizes is a compromise: the performance of small functions suffers from memory transfer delays, but a big function size leads to a badly optimized\footnote{We are not sure that we understand the behavior of the NVidia optimizer. For example, increasing the CUDA kernel size from 2000 arithmetic operations to 3000 ones sometimes slows down the integrand evaluation speed twice.} and slowly compiled code.

We use the techniques for prevention of occasional emergence of very large values that are described in ~\cite{volkov_prd} (with little modifications and adaptation for GPU parallelism).

When we calculate $I(\underline{z})/g(\underline{z})$, it is often the case that machine double precision is not enough for storing $g(\underline{z})$. The machine double precision allows values up to $2^{1025}$. This situation is due to a large number of variables and a closeness of some values of $\ffdeg(s)$ from (\ref{eq_pdf}) to zero. It is not obvious from the beginning that these points can be ignored; see Sec. \ref{sec_res_tech} and Table \ref{table_tech}. To solve this problem, we store $g(\underline{z})$ as $x\cdot 2^j$, where $0.5\leq x<1$ is stored with machine double precision, $j$ is stored as 32-bit integer. 

\subsection{Compilation of the integrands code}

The integrands code was compiled with the NVidia Compiler \texttt{nvcc} into shared libraries that are linked dynamically with the integrator. The compiler is a relatively slow one, and 400 GB of code requires a lot of time for compilation. Like the integration, this compilation was performed on the supercomputer ``Govorun'' from JINR (Dubna, Russia). The processors Intel Xeon Gold 6154 with 18 cores were mostly used for this work. The compilation operation was organized using the MPI protocol with parallel processes that run \texttt{nvcc}: two processes per CPU core. The total compilation time amounted to about 120 CPU-hours.

\subsection{Monte Carlo integration: details}\label{subsec_monte_carlo_details}

The Monte Carlo integrator was written in C++ with CUDA. The integration was performed on several GPUs NVidia Tesla V100 of the supercomputer ``Govorun'' from JINR (Dubna, Russia). Most of the time from 2 to 16 GPUs were occupied for the integration. The inter-device parallelism was organized using the MPI protocol.

The controlling part of the integrator generates the numbers of Feynman graphs to obtain a next block of samples. The number $j$ of a Feynman graph is generated randomly. The probabilities $p_j$ of taking the graph $j$ are chosen to make the convergence as fast as possible. Let us describe the method of obtaining $p_j$. Put
$$
C_j=\sigma_{\uparrow,j} \sqrt{N_j},
$$
where $N_j$ is the number of samples that have already been processed for the graph $j$. By $t_j$ we denote the average time required for evaluation of one integrand value for the graph $j$. The total time that is needed for evaluation of $N$ samples is approximately
$$
t=N\sum_j p_j t_j.
$$
The total standard deviation can be estimated as
$$
\sigma^2=\frac{1}{N} \sum_j \frac{(C_j)^2}{p_j} = \frac{1}{t} \left(\sum_j \frac{(C_j)^2}{p_j}\right)\left( \sum_j p_j t_j\right) = \frac{1}{t} \left(\sum_j \frac{(C_j)^2 t_j}{q_j}\right)\left( \sum_j q_j\right),
$$
where $q_j=p_jt_j$. The minimum point satisfies the equation
$$
\left( \frac{\partial}{\partial q_i} - \frac{\partial}{\partial q_l} \right) \left( \sum_j \frac{(C_j)^2 t_j}{q_j} \right)=0
$$
for any $i,l$. Using this, we obtain
$$
q_j=C C_j\sqrt{t_j},
$$
where $C$ is some constant, or
$$
p_j=\frac{C_j/\sqrt{t_j}}{\sum_l (C_l/\sqrt{t_l})}.
$$
We use this probabilities for random generation of the graph numbers with a little modification for stabilization: a little more attention is being given to the graphs $j$ with big $\sigma_{\uparrow,j}/\sigma_{\downarrow,j}$.

After integration, the total standard deviations (upper and lower) are obtained by
\begin{equation}\label{eq_sigma_sum}
(\sigma_{\uparrow})^2=\sum_j (\sigma_{\uparrow,j})^2,\quad (\sigma_{\downarrow})^2=\sum_j (\sigma_{\downarrow,j})^2.
\end{equation}

\section{RESULTS AND THE TECHNICAL INFORMATION}\label{sec_res_tech}

For reliability, two calculations were performed with different pseudorandom generators, with different choices of the constants $C_2,C_3,C_4$ from (\ref{eq_pdf_final}) and the constant $D$ that is used for defining $g_3$ from (\ref{eq_pdf_final}); see Ref. ~\cite{volkov_gpu}.
\begin{itemize}
\item \texttt{Calc 1}: the generator \texttt{MRG32k3a} from the NVidia CURAND library,
$$
C_2=0.03,\quad C_3=0.035, \quad C_4=0.035,\quad D=0.75.
$$
\item \texttt{Calc 2}: the generator \texttt{Philox\char`_4x32\char`_10} from the NVidia CURAND library,
$$
C_2=0.03,\quad C_3=0.01,\quad C_4=0.06,\quad D=0.75.
$$
\end{itemize}
We use the value
$$
C_1=1-C_2-C_3-C_4
$$
for all calculations.

The calculations have led to the results
$$
A_1^{(10)}[\text{no lepton loops, Calc 1}]=6.74(13),
$$
$$
A_1^{(10)}[\text{no lepton loops, Calc 2}]=6.84(12).
$$
The results were first statistically combined graph-by-graph and then were summed using (\ref{eq_sigma_sum}). These operations are not commutative. Thus, some of the results may look strange\footnote{For example, in Table \ref{table_gauge_5loops} some average values are not in the interval of the source values.}.

The supplemental materials contain the results for all 3213 Feynman graphs for both calculations.

Table \ref{table_gauge_5loops} contains the results for nine gauge-invariant classes $(k,m,m')$ splitting the set of all 5-loop Feynman graphs without lepton loops. By definition, $(k,m,m')$ is the set of all Feynman graphs such that $m$ and $m'$ are the quantities of internal photon lines to the left and to the right from the external photon line (or vice versa), $k$ is the quantity of photons with the ends on the opposite sides of it. In this table, $N_{\text{diag}}$ and $N_{\text{total}}$ are the number of Feynman graphs and the total number of Monte Carlo samples generated for this class.

\tiny
\begin{longtable}{ccccccccc}\caption{Contributions of the gauge invariant classes $(k,m,m')$ to $A^{(10)}_1$; here, $a_i=\int I_i(\underline{z})d\underline{z}$ is the contribution of the $i$-th graph to the value, $I_i$ is the corresponding Feynman parametric integrand. (continued)} \\
\hline \hline Class & Calc 1 & Calc 2 & Value $=\sum_i a_i$ & $\sum_i |a_i|$ & $\max_i |a_i|$ & $\sum_i \int \left|I_i(\underline{z})\right|d\underline{z}$ & $N_{\text{diag}}$ & $N_{\text{total}}$ \\ \hline  \endhead
\caption{Contributions of the gauge invariant classes $(k,m,m')$ to $A^{(10)}_1$; here, $a_i=\int I_i(\underline{z})d\underline{z}$ is the contribution of the $i$-th graph to the value, $I_i$ is the corresponding Feynman parametric integrand.}\label{table_gauge_5loops} \\
\hline \hline Class & Calc 1 & Calc 2 & Value $=\sum_i a_i$ & $\sum_i |a_i|$ & $\max_i |a_i|$ & $\sum_i \int \left|I_i(\underline{z})\right|d\underline{z}$ & $N_{\text{diag}}$ & $N_{\text{total}}$ \\ \hline  \endfirsthead
\hline \endfoot  \hline \hline \endlastfoot
$(1,4,0)$ & 6.158(49) & 6.184(45) & 6.157(33) & 1219.8 & 11.8 & 2521.8 & 706 & $43\times 10^{12}$ \\
$(2,3,0)$ & -0.746(63) & -0.763(59) & -0.754(42) & 3076.8 & 46.2 & 4871.0 & 706 & $73\times 10^{12}$ \\
$(1,3,1)$ & 0.854(50) & 0.972(45) & 0.970(33) & 3170.1 & 67.5 & 3749.9 & 148 & $31\times 10^{12}$ \\
$(3,2,0)$ & -0.399(51) & -0.402(47) & -0.403(34) & 2593.5 & 54.9 & 3783.4 & 558 & $56\times 10^{12}$ \\
$(2,2,1)$ & -2.133(53) & -2.197(50) & -2.165(36) & 3318.1 & 85.0 & 4563.6 & 370 & $48\times 10^{12}$ \\
$(4,1,0)$ & -1.028(31) & -0.991(29) & -1.011(21) & 1199.3 & 56.7 & 1758.2 & 336 & $27\times 10^{12}$ \\
$(1,2,2)$ & 0.312(30) & 0.315(28) & 0.315(20) & 1338.5 & 68.7 & 1515.3 & 55 & $11\times 10^{12}$ \\
$(3,1,1)$ & 2.628(35) & 2.630(33) & 2.625(24) & 1437.3 & 63.5 & 2013.9 & 261 & $26\times 10^{12}$ \\
$(5,0,0)$ & 1.0929(94) & 1.0898(87) & 1.0902(62) & 137.0 & 19.3 & 209.8 & 73 & $39\times 10^{11}$ 
\end{longtable}
\normalsize

\begin{figure}[h]
\begin{center}
\includegraphics{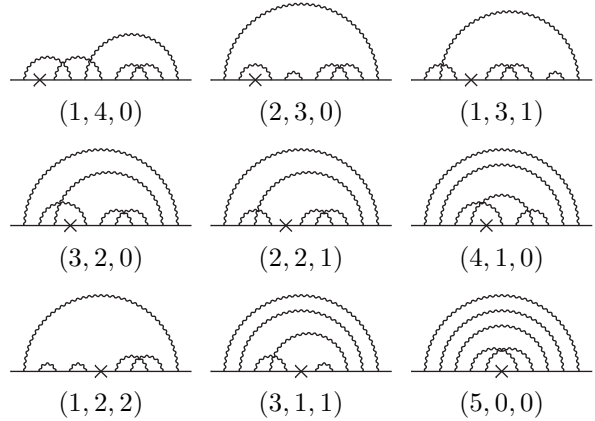}
\end{center}
\caption{Graphs from the gauge-invariant classes $(k,m,n)$ with the maximal absolute values of the contributions.}
\label{fig_max_abs_gauge}
\end{figure}

It was observed by different researchers that the contributions of gauge-invariant classes are relatively small in absolute value, but the contributions of individual Feynman graphs are relatively large and often significantly greater than the class contributions. This occurs regardless of the divergence elimination method used. Table \ref{table_gauge_5loops} demonstrates this fact: the sums and maximums of the graph contribution absolute values are included to the table. Some of the individual graph contributions are 10 times greater than the total contribution. However, this ``oscillating'' nature does not emerge at the level of Feynman parameters. The table demonstrates this too: if the graph contributions are obtained by (\ref{eq_feyn_integral}), then the values of 
$$
\int_{z_1,\ldots,z_M>0}|I(z_1,\ldots,z_M)|\delta(z_1+\ldots+z_M-1) dz_1\ldots
dz_M
$$
are greater than the contribution absolute values only a little; the sums are given in the table. These values are useful for understanding what accuracy can potentially be reached by Monte Carlo integration methods with these integrands. The values for the individual graphs are presented in the supplemental materials. The Feynman graphs with the maximal absolute values of the contributions are presented in FIG. \ref{fig_max_abs_gauge} for each class $(k,m,m')$.

It is very important to check the obtained values independently. However, the amount of computations is huge is this case. Thus, an ability to check the values by parts using different methods would be very useful. We have a splitting of the whole set of graphs into 807 subsets for which the developed subtraction procedure is equivalent to the direct subtraction on the mass shell in Feynman gauge. For each set the equivalence can be proved combinatorially using the Ward identity for individual graphs; see Ref. ~\cite{volkov_gpu}. The splitting is presented in the supplemental materials. It was generated automatically. Each set in this splitting is contained in some gauge-invariant class $(k,m,m')$. There are many sets containing only one graph. The largest set contains 706 graphs: it is the class $(1,4,0)$. We do not know if it is possible to divide this class. An analogous splitting and a comparison with known analytical results is presented in Ref. ~\cite{volkov_2015} for the 3-loop case and in Ref. ~\cite{volkov_gpu} for the 2-loop and 3-loop cases without lepton loops. For the 4-loop case without lepton loops an analogous splitting is presented in Ref. ~\cite{volkov_gpu}, but without a comparison (because no one presented the 4-loop results in the form that is applicable for the comparison). 

\begin{figure}[h]
\begin{center}
\includegraphics{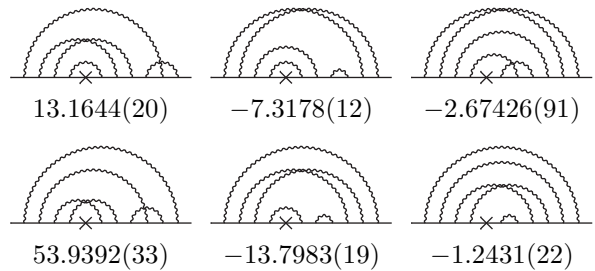}
\end{center}
\caption{The set with the maximum contribution (in absolute value) from the splitting for comparison with the direct subtraction on the mass shell: non-oriented Feynman graphs and their contributions to $A_1^{(10)}$.}
\label{fig_max_abs_split}
\end{figure}

The graph sets from the splitting smooth the peaks of the individual graph contributions as well as the gauge-invariant sets\footnote{It should be noted that this smoothing is not a general principle: for example, the sum of $n$ independent random numbers with the mean values $0$ and the quadratic means $a$ have the quadratic mean $a\cdot\sqrt{n}$.}. However, this ``smoothing'' is not so prominent: some of the set contributions are many times greater than the total contribution (in absolute value). The set with the maximum contribution (in absolute value) is depicted in FIG. \ref{fig_max_abs_split}. This contribution equals $42.0700(50)$.

\small
\begin{longtable}{ccc}\caption{Dependence of the value and the estimated error on the number of Monte Carlo samples $N_{\text{total}}$: $A_1^{(10)}[\text{no lepton loops}]$, Calc 2 (continued)} \\
\hline \hline $N_{\text{total}}$ & Value & $\sigma_{\uparrow}/\sigma_{\downarrow}$ \\ \hline  \endhead
\caption{Dependence of the value and the estimated error on the number of Monte Carlo samples $N_{\text{total}}$: $A_1^{(10)}[\text{no lepton loops}]$, Calc 2}\label{table_dependence_5loops} \\
\hline \hline $N_{\text{total}}$ & Value & $\sigma_{\uparrow}/\sigma_{\downarrow}$ \\ \hline  \endfirsthead
\hline \endfoot  \hline \hline \endlastfoot
$5\times 10^{11}$ & 9(13) & $2.40$ \\
$10^{12}$ & 10.2(8.9) & $2.45$ \\
$2\times 10^{12}$ & 11.2(5.4) & $2.42$ \\
$5\times 10^{12}$ & 9.4(2.6) & $2.25$ \\
$10^{13}$ & 7.9(1.4) & $2.10$ \\
$2\times 10^{13}$ & 7.21(53) & $1.67$ \\
$5\times 10^{13}$ & 6.88(24) & $1.38$ \\
$10^{14}$ & 6.80(16) & $1.34$ \\
$17\times 10^{13}$ & 6.84(12) & $1.31$ 
\end{longtable}
\normalsize

Table \ref{table_dependence_5loops} contains the dependence of the total calculated value and the error on the number of Monte Carlo samples for \texttt{Calc 2}.

Table \ref{table_tech} contains some technical information about the calculations \texttt{Calc 1} and \texttt{Calc 2}. The fields of the table have the following meaning:
\begin{itemize}
\item Value is the obtained value for $A_1^{(10)}[\text{no lepton loops}]$
 with the uncertainty $\sigma_{\uparrow}$; see Sec. \ref{subsec_carlo_values} and Ref. ~\cite{volkov_gpu};
\item $\sigma_{\uparrow}/\sigma_{\downarrow}$ is the relation
between the improved standard deviation and the conventional one,
see Sec. \ref{subsec_carlo_values} and Ref. ~\cite{volkov_gpu};
\item $N_{\text{total}}$ is the total quantity of Monte Carlo samples;
\item $N^{\text{fail}}_{\text{EIA}}$ is the quantity of samples for
which eliminated interval arithmetic failed; see Sec. \ref{subsec_realization_integrands} and Ref. ~\cite{volkov_gpu}; 
\item $\triangle^{\text{fail}}_{\text{EIA}}$ is the contribution of
that samples;
\item $N^{\text{fail}}_{\text{IA}}$ is the quantity of samples for
which direct double-precision interval arithmetic failed;
\item $\triangle^{\text{fail}}_{\text{IA}}$ is the contribution of
that samples;
\item $N^{\text{fail}}_{\text{128}}$, $N^{\text{fail}}_{\text{192}}$, $N^{\text{fail}}_{\text{256}}$ are the quantities of samples for
which the interval arithmetic based on numbers with 128-bit, 192-bit, 256-bit mantissa
failed;
\item $\triangle^{\text{fail}}_{\text{128}}$, $\triangle^{\text{fail}}_{\text{192}}$ are the contributions of that samples;
\item $N^{\text{dens}}_{\text{out of double}}$ is the quantity of samples for
which machine double precision was not enough for storing the probability density; see Sec. \ref{subsec_realization_integrands};
\item $\triangle^{\text{dens}}_{\text{out of double}}$ is the contribution of
that samples;
\item GFlops = billions floating point number operations per second (during the evaluation of the integrands); GIntervals  = billions interval operations per second (in the sense of interval arithmetic); M = millions.
\end{itemize}
\small
\begin{longtable}{lcc}\caption{Technical information about the calculations (continued)} \\
\hline \hline  & Calc 1 & Calc 2 \\ \hline  \endhead
\caption{Technical information about the calculations}\label{table_tech} \\
\hline \hline  & Calc 1 & Calc 2 \\ \hline  \endfirsthead
\hline \endfoot  \hline \hline \endlastfoot
Value & $6.74(13)$ & $6.84(12)$ \\
$\sigma_{\uparrow}/\sigma_{\downarrow}$ & $1.31$ & $1.31$ \\
$N_{\text{total}}$ & $15\times 10^{13}$ & $17\times 10^{13}$ \\
$N^{\text{fail}}_{\text{EIA}}$ & $34\times 10^{12}$ & $39\times 10^{12}$ \\
$N^{\text{fail}}_{\text{IA}}$ & $38\times 10^{10}$ & $42\times 10^{10}$ \\
$N^{\text{fail}}_{\text{128}}$ & $67\times 10^{6}$ & $73\times 10^{6}$ \\
$N^{\text{fail}}_{\text{192}}$ & $10787$ & $2453$ \\
$N^{\text{fail}}_{\text{256}}$ & $8669$ & $0$ \\
$N^{\text{dens}}_{\text{out of double}}$ & $11\times 10^{5}$ & $13\times 10^{5}$ \\
$\triangle^{\text{fail}}_{\text{EIA}}$ & $4$ & $5$ \\
$\triangle^{\text{fail}}_{\text{IA}}$ & $0.9$ & $3$ \\
$\triangle^{\text{fail}}_{\text{128}}$ & $-0.07$ & $-0.07$ \\
$\triangle^{\text{fail}}_{\text{192}}$ & $-0.002$ & $-3\times 10^{-6}$ \\
$\triangle^{\text{dens}}_{\text{out of double}}$ & $-6\times 10^{-13}$ & $6\times 10^{-10}$ \\
Total calculation time, GPU-hours & 19515 & 20341 \\
Share in the time: double-precision EIA & $21.6\%$ & $23.3\%$ \\
Share in the time: double-precision IA & $35.5\%$ & $34.5\%$ \\
Share in the time: 128-bit-mantissa IA & $28.1\%$ & $29.1\%$ \\
Share in the time: 192-bit and 256-bit-mantissa IA & $11.4\%$ & $10.0\%$ \\
Share in the time: sample generation & $1.8\%$ & $1.5\%$ \\
Share in the time: other operations & $1.7\%$ & $1.7\%$ \\
GPU speed: double-precision EIA, GFlop/s & $2221.88$ & $2227.99$ \\
GPU speed: double-precision EIA, GInterval/s & $1962.13$ & $1965.60$ \\
GPU speed: double-precision IA, GFlop/s & $1358.63$ & $1505.30$ \\
GPU speed: double-precision IA, GInterval/s & $274.13$ & $303.31$ \\
GPU speed: 128-bit-mantissa IA, GFlop/s & $13.47$ & $13.48$ \\
GPU speed: 128-bit-mantissa IA, GInterval/s & $2.54$ & $2.53$ \\
GPU speed: 192-bit and 256-bit-mantissa IA, MFlop/s & $4.21$ & $4.65$ \\
GPU speed: 192-bit and 256-bit-mantissa IA, MInterval/s & $0.74$ & $0.80$ 
\end{longtable}
\normalsize

It is easy to see that in EIA one arithmetic operation on intervals takes approximately one operation on numbers. This is due to the fact that the most part of the EIA calculation is occupied by the operations on the centers of the intervals. However, in IA one interval operation takes approximately five operations on numbers. Also, the speed of the number operations for IA is by 1.6 times less than for EIA. This is because most of the operations in IA require specifying a rounding mode\footnote{However, this difference in the speed was not discovered in the calculations on NVidia Tesla K80 from Ref. ~\cite{volkov_gpu} despite the fact that the difference was discovered during the preliminary tests.}, but the operations on the centers of intervals in EIA do not require it.

\texttt{Calc 1} suffered from some errors that cause an emergence of anomalous points that have contributions to $N^{\text{fail}}_{\text{192}}$, $N^{\text{fail}}_{\text{256}}$, $N^{\text{dens}}_{\text{out of double}}$; see Table \ref{table_tech}. We can not perform the full recalculation because this requires a lot of time. However, that points do not have a significant impact on the results; the table confirms this fact. That errors were corrected in \texttt{Calc 2}.


Table \ref{table_tech} demonstrates that the points requiring an increased precision have a significant contribution to the result. For example, $\triangle^{\text{fail}}_{\text{EIA}}$ and 
$\triangle^{\text{fail}}_{\text{IA}}$ are at the level of the total contribution, $\triangle^{\text{fail}}_{\text{128}}$ is at the level of the uncertainty. Also, the table shows that that contributions are unstable due to an ``oscillating'' character of the individual graph contributions, a floating character of the interval acception criteria (\ref{eq_interval_accept}), and a difference in the probability density functions. In addition, the table shows that the contribution $\triangle^{\text{dens}}_{\text{out of double}}$ is insignificant. However, this contribution is too far from the boundaries of machine double precision like $2^{-1025}$ (on a logarithmic scale). Thus, there may be situations, where such contributions will be significant. This fact demonstrates that universal Monte Carlo integration routines can work poorly for many-loop Feynman parametric integrals.

\begin{figure}[h]
\begin{center}
\includegraphics[scale=0.5]{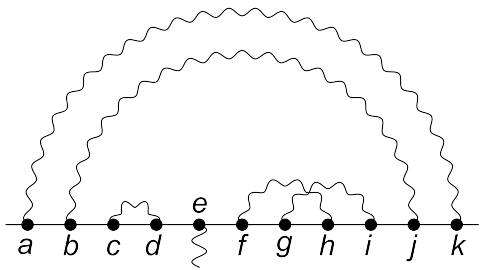}
\end{center}
\caption{The graph with the maximum (in absolute value) contribution of the Monte Carlo samples for which eliminated interval arithmetic failed.}
\label{fig_graph_eia_fail}
\end{figure}

\begin{figure}[h]
\begin{center}
\includegraphics{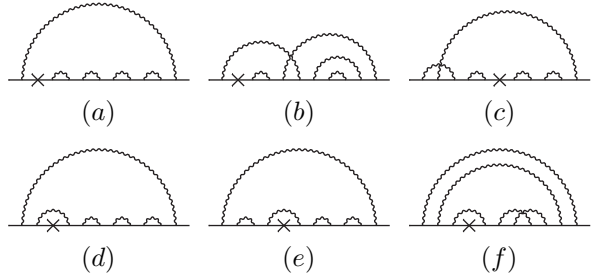}
\end{center}
\caption{The extreme graphs of different kinds: (a) best Monte Carlo integration convergence quality; (b) worst Monte Carlo integration convergence quality; (c,d,e) maximal (in absolute value) contribution of the samples for which the interval arithmetic with numbers of double precision, 128-bit mantissa, 192-bit mantissa failed; (f) maximal (in absolute value) contribution of the samples for which double precision was not enough for storing the probability density.}
\label{fig_extr_graphs}
\end{figure}

An analogous information for the individual Feynman graphs is contained in the supplemental materials. The graphs with the maximal contributions to $\triangle^{\text{fail}}_{\text{EIA}}$, $\triangle^{\text{fail}}_{\text{IA}}$, $\triangle^{\text{fail}}_{\text{128}}$, $\triangle^{\text{fail}}_{\text{192}}$, $\triangle^{\text{dens}}_{\text{out of double}}$ are shown in FIG. \ref{fig_graph_eia_fail} and FIG. \ref{fig_extr_graphs} (c--f). The corresponding contributions (for \texttt{Calc 2}) are 
$$
67.1,\quad 26.3,\quad 0.15,\quad 3.1\cdot 10^{-5},\quad 5.9\cdot 10^{-10}.
$$
The Monte Carlo integration convergence quality for a given graph $j$ can be estimated as
$$
\frac{\sigma_{\uparrow,j}\cdot \sqrt{N_j}}{\int \left|I_j(\underline{z})\right|d\underline{z}},
$$
where $N_j$ is the number of Monte Carlo samples for the $j$-th graph, $I_j$ is the corresponding Feynman parametric integrand. Less values correspond to a better quality. The graphs with the best and the worst quality are shown in FIG. \ref{fig_extr_graphs} (a,b). The corresponding values (for \texttt{Calc 2}) are
$$
16.2,\quad 525.9.
$$
These values demonstrate that even in the best case the Monte Carlo integration works not ideally due to large dimensionality. However, this is acceptable and requires a relatively small amount of the supercomputer time for integration.

\section{CONCLUSION}

A numerical calculation of the total contribution of the 5-loop QED Feynman graphs without lepton loops to the corresponding coefficient of the electron anomalous magnetic moment expansion in $\alpha$ was performed. The calculation is based on a specific method of reduction of the problem to Feynman parametric integrals and on Monte Carlo integration using a supercomputer. Usage of some mathematical considerations about the integrands behavior provided us an ability to reduce the amount of the needed supercomputer power and time significantly. 

This calculation provides the first independent check of the value obtained by T. Kinoshita's team that is presented in Ref. ~\cite{kinoshita_atoms}. However, the discrepancy of about $4.8\sigma$ between the results was discovered. On the one hand, this discrepancy does not significantly affect the known values of $a_e$ and $\alpha$. But on the other hand, it requires an additional independent calculation and can affect the physics in the future.

The results of the calculation are presented in detail. This detailed presentation gives us an ability to check the results by parts using another methods. The contribution values of nine gauge-invariant classes splitting the whole set are presented for the first time (except the preliminary values in Ref. ~\cite{volkov_acat}).

For reliability, two different Monte Carlo integrations with different pseudorandom generators were performed. The results of these calculations agree with each other, and they were stastistically combined in the final result. 

A cancellation of an ``oscillating'' nature of the individual Feynman graph contributions in the gauge-invariant classes confirms that the results are correct. This ``oscillating'' nature is described in detail. However, there is no mathematical foundation for this cancellation at the current moment of time. Also, it is surprising that we have only an inter-graph oscillation, but not in Feynman parametric space for one graph. 

The technical information that is presented in the paper will be useful for the scientists that are going to perform many-loop calculations in quantum field theory or another computations using supercomputers and graphics accelerators. Also, the provided information about the Monte Carlo integration will be useful for developers of Monte Carlo integrators.

In closing, let us recapitulate some problems that still
remain open:
\begin{enumerate}
\item To perform an independent calculation of the 5-loop contribution of the graphs with lepton loops; to check the value from Ref. ~\cite{kinoshita_10_corr}.
\item To prove rigorously (or disprove) that the developed subtraction procedure (Ref. ~\cite{volkov_2015}) leads to finite integrals for each suitable Feynman graph;
\item To substantiate rigorously the developed Monte Carlo integration method (Ref. ~\cite{volkov_prd}) and to extend it to the graphs with lepton loops;
\item To explain why the ``oscillating'' nature of the individual Feynman graph contributions is cancelled in the gauge-invariant classes.
\end{enumerate}

\section*{ACKNOWLEDGEMENTS}

The author thanks Andrey Kataev for helpful recommendations, Lidia Kalinovskaya for her help in organizational issues, and Predrag Cvitanovi\'{c} for the ideas about gauge-invariant classes. Also, the author thanks the Laboratory of Information Technologies of JINR (Dubna, Russia) for providing an access to the supercomputer ``Govorun'' and the organizers of the conference ACAT-2019 (Saas Fee, Switzerland, March 2019) for providing an ability to present the preliminary results at the conference without financial problems.

\end{document}